\documentclass[twocolumn,showpacs,preprintnumbers,amsmath,amssymb,prl]{revtex4}
\usepackage{amsmath, amssymb}

\usepackage{graphicx}
\usepackage{dcolumn}
\usepackage{bm}

\newcommand{\Fig}[1]{Fig.\,\ref{fig:#1}}
\newcommand{\fig}[1]{Fig.\,\ref{fig:#1}}

\newcommand{\eq}[1]{Eq.\,(\ref{eq:#1})}
\newcommand{\PPKTP}{periodically poled KTiOPO$_4$\,}
\newcommand{\scite}[1]{\,\cite{#1}}

\begin{document}

\title{Observation of -9~dB quadrature squeezing with improvement of phase stability in homodyne measurement}

\author{
	Yuishi Takeno, Mitsuyoshi Yukawa, Hidehiro Yonezawa, and Akira Furusawa
}

\address{
	Department of Applied Physics, School of Engineering, The University of Tokyo,\\ 7-3-1 Hongo, Bunkyo-ku, Tokyo 113-8656, Japan \\
	CREST, Japan Science and Technology Agency, 1-9-9 Yaesu, Chuo-ku, Tokyo 103-0028, Japan
}

\begin{abstract}
We observe -9.01$\pm$0.14~dB of squeezing and +15.12$\pm$0.14~dB of antisqueezing with a local oscillator phase locked in homodyne measurement.
In reference \cite{SYKSF.APL.2006}, two main factors are pointed out which degrade the observed squeezing level:
	phase fluctuation in homodyne measurement
	and intracavity losses of an optical parametric oscillator for squeezing.
We have improved the phase stability of homodyne measurement and have reduced the intracavity losses. 
We measure pump power dependences of the squeezing and antisqueezing levels, which show good agreement with theoretical calculations taking account of the phase fluctuation.
\end{abstract}


\maketitle



\section{Introduction}
Squeezed states can be important resources for quantum information processing with continuous variables\scite{BL.RMP.2005}.
For example, quadrature squeezed vacuum states are used to realize quantum teleportation\scite{FSBFKP.SCI.1998} which is one of the most important protocols in quantum information processing.
The squeezing level limits the fidelity of teleportation\scite{FSBFKP.SCI.1998, BFKL.PRA.2001}.
Hence it is important to improve the squeezing level to achieve higher fidelity.

A quadrature squeezed vacuum state of continuous-wave (cw) light is often generated by utilizing a subthreshold optical parametric oscillator (OPO), and a \PPKTP (PPKTP) crystal is being used as a nonlinear optical medium in the OPO\scite{ATF.OEX.2006, SYKSF.APL.2006, TAYFK.OLE.2006, HGPHBBL.JPB.2007}.
\text{-7.2$\pm$0.2~dB} of squeezing at 860~nm has been reported by Suzuki et al.\scite{SYKSF.APL.2006} using a PPKTP crystal.
As is indicated in this reference, one is required to suppress fluctuation of relative phase between squeezed light and a local oscillator (LO) beam in homodyne measurement and to reduce intracavity losses of the OPO in order to observe higher level of squeezing.
Therefore we have focused on suppression of the phase fluctuation and reduction of the intracavity losses to improve the observed squeezing level.

\section{Experiment}

\begin{figure}[tbh]
\begin{center}
\includegraphics[width=0.8\linewidth]{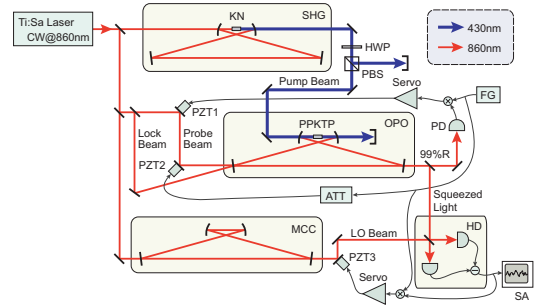}
\caption{
	(Color online)
	Schematic of experimental setup.
	SHG: second harmonic generator,
	KN: KNbO$_3$ crystal,
	OPO: optical parametric oscillator,
	PPKTP: \PPKTP crystal,
	PD: photo detector,
	MCC: mode cleaning cavity,
	LO: local oscillator,
	HD: balanced homodyne detector,
	$\ominus$: subtraction circuit,
	$\otimes$: mixing circuit (multiplier),
	FG: function generator,
	ATT: electrical attenuator,
	Servo: servo amplifier circuit for feedback system,
	PZTs: piezoelectric transducers,
	and SA: spectrum analyzer.
}
\label{fig:exp}
\end{center}
\end{figure}

\Fig{exp} shows a schematic of our experimental setup.
A quadrature squeezed vacuum state is generated from a subthreshold OPO which is pumped by a second harmonic light generated with doubling of an output of a cw Ti:Sapphire laser (Coherent, MBR110) at 860~nm.
A cavity of the OPO has a bow-tie-type ring configuration consisting of two spherical mirrors (radius of curvature of 50~mm) and two flat mirrors.
One of the two flat mirrors is an output coupler and the transmittance is $T=0.123$.
The round trip length is about 500~mm.
A 10-mm-long PPKTP crystal (Raicol Crystals) is placed between the two spherical mirrors.
The PPKTP crystal is specially antireflecting-coated to reduce the intracavity losses as much as possible.
The beam waist size inside the crystal is about 20~$\mu$m.

Squeezed light from the OPO is measured by a balanced homodyne detector and the output electric signal is analyzed by a spectrum analyzer (Agilent, E4402B).
The homodyne detector consists of two Si photodiodes (Hamamatsu, S3590-06 with antirefrection coating) and a subtraction circuit.
At the homodyne detection, we need to lock the relative phase between the squeezed light and the LO beam in order to measure the squeezed or antisqueezed quadrature.
To realized this phase locking, we use a phase-modulated weak coherent beam, which is called a ``probe" beam, and the phase modulation is applied to it by a function generator (FG) and a piezoelectric transducer (PZT2).
At first, we lock a relative phase between the pump beam and the probe beam by monitoring the intensity of the transmitted probe beam from the OPO with a photo detector (PD) to make the probe beam intensity minimum or maximum by controlling the PZT1, i.e., the parametric gain is locked at the most de-amplified phase or the most amplified phase.
Then, we lock a relative phase between the LO beam and the probe beam by using the output signal of the homodyne detector to get an error signal for controlling the PZT3.
As a result of these two phase locking, the relative phase between the LO beam and the squeezed light is locked so that we can measure the squeezed or antisqueezed quadrature.

The relative phase is fluctuating even if it is locked.
Because of the finite phase fluctuation, the antisqueezed quadrature contaminates the observed squeezing level.
In order to observe high level of squeezing, it is essentially required to suppress the phase fluctuation.
Taking account of the phase fluctuation with an rms of $\tilde{\theta}$, the observed squeezing level $R^\prime_{-}$ and antisqueezing level $R^\prime_{+}$ are calculated as follows\scite{PCK.APB.1992, ZGCLK.PRA.2003, ATF.OEX.2006}:
\begin{gather}
R_{\pm}^\prime \approx R_{\pm}\cos^2\tilde{\theta}+R_{\mp}\sin^2\tilde{\theta}, \label{eq:pf}\\
R_{\pm} = 1\pm\eta\xi^2\zeta\rho\frac{4x}{\left(1\mp x\right)^2+4\Omega^2} \label{eq:sl},
\end{gather}
where $R_{\pm}$ is generated squeezing/antisqueezing levels without taking account of the phase fluctuation, $\eta$ is quantum efficiency of the photodiode, $\xi$ is homodyne visibility, $\zeta$ is propagation efficiency, $\rho = T/(T+L)$ is escape efficiency of the cavity, $T$ is transmittance of the output coupler, $L$ is the intracavity losses, $x = \sqrt{P/P_\mathrm{th}}$ is normalized pump power, $P$ is pump power, $P_\mathrm{th}$ is an oscillation threshold of the OPO, $\Omega = f/\gamma$ is normalized frequency, $f$ is measurement frequency by a spectrum analyzer, $\gamma = c(T+L)/l$ is the OPO cavity decay rate, and $l$ is round trip length of the cavity.

\section{Improvement}

One of the main factors we need to improve is the phase fluctuation.
In the reference \cite{SYKSF.APL.2006}, the phase fluctuation $\tilde{\theta}$ was 3.9$^\circ$, which limits the observed squeezing level to -8.66~dB even in an ideal, lossless condition.
To reduce the phase fluctuation, we have improved our feedback system, especially the following two points:
	resonant frequency of the feedback system,
	and signal-to-noise ratio (SNR) of the error signals.

To make the resonant frequency of the feedback system higher, we have replaced PZTs from single-layer ones to stuck ones (Thorlabs, AE0203D04).
Generally speaking, a stuck PZT has larger capacitance than a single-layer one so that it seems that the feedback system has lower resonant frequency with a stuck PZT.
But resonance of the feedback system in fact occurs at an eigen frequency of a complex consisting of a PZT, a mirror, and a mount, and the eigen frequency of the system is lower than the frequency derived from only an effect of the capacitance.
In our case, the resonant frequency was about 2~kHz with single-layer PZTs, and with the present setup the resonance occurs at 14~kHz at PZT1 for parametric gain locking and at 22~kHz at PZT3 for LO phase locking, respectively.
Note that the difference between these two resonant frequencies would be caused by individualities of the PZTs.
As a result of the higher resonant frequencies, we can apply larger feedback gain to low frequency domain to stabilize the phase locking.

To increase SNR of the error signals, we have optimized modulation frequency, modulation depth, and probe beam power.
The modulation frequency should be sufficiently high not to be affected by original laser noise around 89~kHz and 177~kHz.
Thus we have set it 295~kHz because the laser noise can be considered at the shot noise level and the PZT2 resonates at this frequency.
The modulation depth and the probe beam power should be large enough to obtain clear error signals, but they can affect homodyne measurement harmfully.
We have optimized them and checked that the shot noise level was not affected by the modulation at the measurement frequency.

Thanks to these improvements, the phase fluctuation is suppressed to 1.5$^\circ$, with which the highest squeezing level is supposed to be -12.81~dB in an ideal, lossless condition of the present experimental setup.

Another main factor which degrades the observed squeezing level is the intracavity losses.
In the reference \cite{SYKSF.APL.2006}, intracavity losses were $L = 0.0063$ which stayed constant independently of the pump beam power, that is, blue light induced infrared absorption (BLIIRA)\scite{WPL.JAP.2004, MPK.JOSAB.1994} or other pump light induced losses were not observed.
To reduce the intracavity losses as much as possible, we use a specially antireflecting-coated PPKTP crystal.
With the crystal, the intracavity losses are reduced to $L = 0.0020$ without a pump beam and $L = 0.0038$ with a pump beam at the power of 100~mW.
Therefore the escape efficiency $\rho = T/(T+L)$ is improved from 0.95 to 0.97 under 100~mW pumping.
Note that small pump light induced losses are observed with the new crystal, and the induced losses increase as the pump beam power increases.
Moreover, if we make the wavelength of the output beam of the Ti:Sa laser slightly shorter ($\Delta\lambda\approx$-1~nm), the intracavity losses become to grow gradually in time (the intracavity losses increased to more than 0.0073 at 2000 seconds after 100~mW pump beam was injected).
This phenomenon depends strongly on the wavelength so that we have chosen better wavelength without the effect.

\section{Results and discussions}

\begin{figure}[tbh]
\begin{center}
\includegraphics[width=0.8\linewidth]{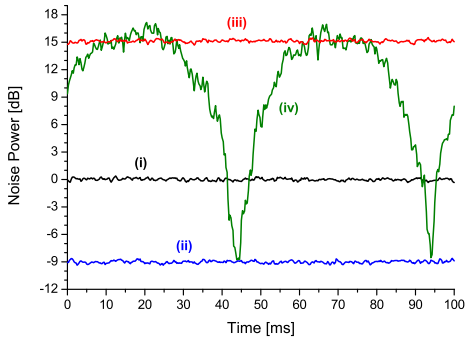}
\caption{
	(Color online)
	Observed noise levels at the pump power of 100~mW.
	(i) Shot noise level.
	(ii) LO phase is locked at the squeezed quadrature.
	(iii) LO phase is locked at the antisqueezed quadrature.
	(iv) LO phase is scanned.
	These levels are normalized to make the shot noise level 0~dB.
	Measurement frequency: 1~MHz,
	resolution bandwidth: 30~kHz,
	and video bandwidth: 300~Hz.
	Traces (i), (ii), and (iii) are averaged for 20 times.
	The observed squeezing/antisqueezing levels are -9.01$\pm$0.14/+15.12$\pm$0.14~dB, respectively.
}
\label{fig:sq}
\end{center}
\end{figure}

\Fig{sq} shows the observed noise levels at the pump power of 100~mW.
All traces are measured by a spectrum analyzer with the following conditions:
	measurement frequency is 1~MHz in a zero span mode,
	resolution bandwidth is 30~kHz,
	and video bandwidth is 300~Hz.
The trace (i) is the shot noise level, (ii) a noise level with the LO phase locked at the squeezed quadrature, (iii) locked at the antisqueezed quadrature, and (iv) a noise level with the LO phase scanned.
These traces except for (iv) are averaged for 20 times.
The observed squeezing level is -9.01$\pm$0.14~dB and the antisqueezing level is +15.12$\pm$0.14~dB.

In this measurement, the intracavity losses are $L = 0.0038$ with 100~mW pump beam, the parametric gain (amplification factor) $G = 18.7$, propagation efficiency $\zeta = 0.99$, the phase fluctuation $\tilde{\theta} = 1.5^\circ$, the homodyne visibility $\xi = 0.988$, photodiodes quantum efficiency $\eta = 0.998$, and the detector circuit noise level is -21.7~dB compared to the shot noise level at the LO power of 5.5~mW.
Subtracting the detector circuit noise, the squeezing/antisqueezing levels $R_{\pm}^\prime$ are -9.22$\pm$0.15/+15.15$\pm$0.14~dB, respectively.
Considering \eq{pf}, the generated squeezing/antisqueezing levels $R_{\pm}$ are estimated as -10.12$\pm$0.18/+15.15$\pm$0.14~dB, respectively.
Here, the total losses can be estimated as 0.0709$\pm$0.0045 via \eq{sl}, while by using measurement values the total losses $1-\rho\zeta\xi^2\eta = 0.0645$, which agree well with the estimation.

\begin{figure}[tbh]
\begin{center}
\includegraphics[width=0.8\linewidth]{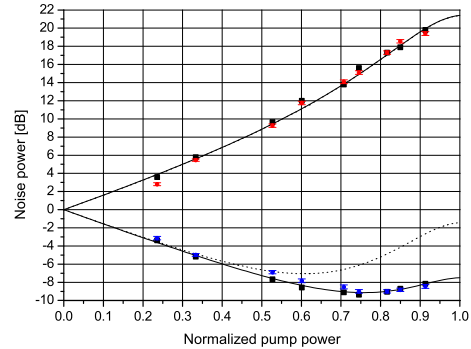}
\caption{
	(Color online)
	Pump power dependences of the squeezing and antisqueezing levels.
	Normalized pump power is calculated by $\sqrt{P/P_\mathrm{th}}$ where $P$ is the pump power and $P_\mathrm{th} = 180$~mW is the oscillation threshold of the OPO.
	Blue (red) circles indicate the observed squeezing (antisqueezing) levels.
	Black squares and solid/dotted curves indicate theoretically calculated values.
	Each square plot is calculated by using the measured intracavity losses at each pump power while the curves are calculated by using \eq{fil} in the condition of the phase fluctuation $\tilde{\theta} = 1.5^\circ$ (solid) and $3.9^\circ$ (dotted).
}
\label{fig:ppd}
\end{center}
\end{figure}

We repeat the above measurement for several pump powers.
The observed squeezing and antisqueezing levels are shown in \fig{ppd}.
In the figure, theoretical values are derived from \eq{pf} using the measured values in the experiment.
Black squares are calculated by the measured intracavity losses for each pump powers, and traces (solid and dotted) are calculated by using the following equation:
\begin{equation}
L(x) = 0.00249+0.00222 x. \label{eq:fil}
\end{equation}
This equation is derived by fitting the measured intracavity losses for several pump powers.
The solid curves are calculated with $\tilde{\theta} = 1.5^\circ$, and the dotted with 3.9$^\circ$\scite{SYKSF.APL.2006}.
The horizontal axis represents normalized pump power $x = \sqrt{P/P_\mathrm{th}}$ which is more convenient than pump power $P$ because the subthreshold condition is satisfied within the range of $0\leq P<P_\mathrm{th}$, corresponding to $0\leq x<1$.
The measurements are performed at the pump power of 10, 20, 50, 65, 90, 100, 120, 130, and 150~mW.
To calculate the normalized pump powers from these ``raw" pump powers, we derive oscillation threshold $P_\mathrm{th}$ of the OPO by the following relations\scite{ZGCLK.PRA.2003}:
\begin{gather}
G(P) = \frac{1}{\left(1-x(P)\right)^2}, \\
x(P) = \sqrt{\frac{P}{P_\mathrm{th}}}.
\end{gather}
By using these equations, the oscillation threshold is estimated as $P_\mathrm{th} = 180\pm 2$~mW using the measured parametric gains $G(P)$ for several pump powers $P$.

The theoretically calculated values show good agreement with the experimental results, which suggests that \eq{pf} is a proper model of degradation of squeezing caused by phase fluctuation.
One can see that the observed squeezing level degrades above $x \approx 0.82$ since the large antisqueezed quadrature contaminates the observed (squeezed) quadrature due to the finite phase fluctuation so that there exists optimal normalized pump power $x_\mathrm{opt}$ to observe the highest squeezing level.
Of course the optimal power $x_\mathrm{opt}$ increases as the phase fluctuation $\tilde{\theta}$ is suppressed.
Thus it is obvious that suppression of the phase fluctuation is extremely effective to observe high level of squeezing.

\begin{figure}[tbh]
\begin{center}
\includegraphics[width=0.8\linewidth]{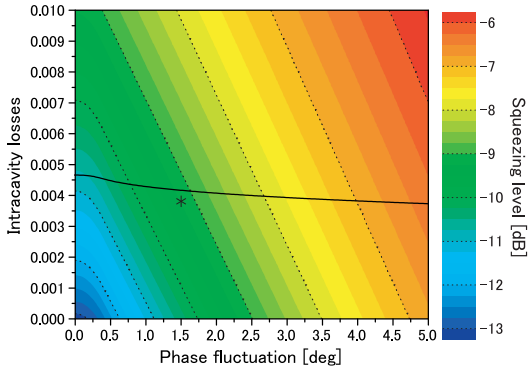}
\caption{
	(Color online)
	Theoretically predicted squeezing level $R_{-}^\prime(\tilde{\theta}, L, x_\mathrm{opt}(\tilde{\theta}, L))$.
	Solid curve satisfies \eq{fil}, which represents the relation between $x$ and $L$ in the present experimental condition.
	Our current position ($\tilde{\theta}$, $L$) = (1.5$^\circ$, 0.0038) is shown by $\ast$.
}
\label{fig:lev}
\end{center}
\end{figure}

\Fig{lev} shows the theoretically predicted squeezing level $R_-^\prime(\tilde{\theta}, L, x_\mathrm{opt})$ taking account of the phase fluctuation and also detector's circuit noise.
In the calculation, experimentally obtained values are used except for the phase fluctuation $\tilde{\theta}$, the intracavity losses $L$, and the normalized pump power $x$.
The axis of $x$ is reduced by choosing the optimal value $x_\mathrm{opt}(\tilde{\theta}, L)$ mentioned above.
A solid curve satisfies \eq{fil}, which represents the relation between $x$ and $L$ in the present experimental condition.
We can advance along the curve only by improving the phase stability in homodyne measurement.
We are now at $(\tilde{\theta}, L) = (1.5^\circ, 0.0038)$, represented by $\ast$, and have already exceeded the -9~dB line in the figure.
With much more stable system ($\tilde{\theta} \approx 0^\circ$), it would be possible to observe a squeezing level over -10~dB.
This figure can indicate the future prospects of the squeezing level, having two parameters $\tilde{\theta}$ and $L$ which are the main factors degrading the observed squeezing level.

\section{Conclusion}

Thanks to suppression of the phase fluctuation and reduction of the intracavity losses, we succeed in observing -9.01$\pm$0.14~dB of squeezing and +15.12$\pm$0.14~dB of antisqueezing with the LO phase locked.
Theoretical calculation shows good agreement with the experimental results for several pump powers and we observe the degradation of squeezing level with sufficiently large pump powers as predicted by the model of the phase fluctuation.
In order to achieve higher level of squeezing, it is extremely important to obtain a more stable system.

\section*{Acknowledgements}

One of the authors (Y.T.) profoundly thanks Dr. Shigenari Suzuki.
This work was partly supported by the MEXT of Japan.


\begin{thebibliography}{99}
	\bibitem{SYKSF.APL.2006}
	S. Suzuki, H. Yonezawa, F. Kannari, M. Sasaki, and A. Furusawa,
	"7~dB quadrature squeezing at 860~nm with periodically poled KTiOPO$_4$",
	Appl. Phys. Lett. {\bf 89}, 061116 (2006).

	\bibitem{BL.RMP.2005}
	S. L. Braunstein and Peter van Loock,
	"Quantum information with continuous variables",
	Rev. Mod. Phys. {\bf 77}, 513-577 (2005).

	\bibitem{FSBFKP.SCI.1998}
	A. Furusawa, J. L. S\o rensen, S. L. Braunstein, C. A. Fuchs, H. J. Kimble, and E. S. Polzik,
	"Unconditional Quantum Teleportation",
	Science {\bf 282}, 706-709 (1998).

	\bibitem{BFKL.PRA.2001}
	S. L. Braunstein, C. A. Fuchs, H. J. Kimble, and P. van Loock,
	"Quantum versus classical domains for teleportation with continuous variables",
	Phys. Rev. A {\bf 64}, 022321 (2001).

	\bibitem{ATF.OEX.2006}
	T. Aoki, G. Takahashi, and A. Furusawa,
	"Squeezing at 946nm with periodically poled KTiOPO$_4$",
	Opt. Exp. {\bf 14}, 6930-6935 (2006).

	\bibitem{TAYFK.OLE.2006}
	T. Tanimura, D. Akamatsu, Y. Yokoi, A. Furusawa, and M. Kozuma,
	"Generation of a squeezed vacuum resonant on a rubidium D$_1$ line with periodically poled KTiOPO$_4$",
	Opt. Lett. {\bf 31}, 2344-2346 (2006).

	\bibitem{HGPHBBL.JPB.2007}
	G. H\'{e}tet, O. Gl\"{o}ckl, K. A. Pilypas, C. C. Harb, B. C. Buchler, H. A. Bachor, and P. K. Lam,
	"Squeezed light for bandwidth-limited atom optics experiments at the rubidium D1 line",
	J. Phys. B {\bf 40} 221-226 (2007).

	\bibitem{PCK.APB.1992}
	E. S. Polzik, J. Carri, and H. J. Kimble,
	"Atomic Spectroscopy with Squeezed Light for Sensitivity Beyond the Vacuum-State Limit",
	Phys. Rev. B {\bf 55}, 279-290 (1992).

	\bibitem{ZGCLK.PRA.2003}
	T. C. Zhang, K. W. Goh, C. W. Chou, P. Lodahl, and H. J. Kimble,
	"Quantum teleportation of light beams",
	Phys. Rev. A {\bf 67}, 033802 (2003).

	\bibitem{WPL.JAP.2004}
	S. Wang, V. Pasiskevicius, and F. Laurell,
	"Dynamics of green light-induced infrared absorption in KTiOPO$_4$ and periodically poled KTiOPO$_4$",
	J. Appl. Phys. {\bf 96}, 2023-2028 (2004).

	\bibitem{MPK.JOSAB.1994}
	H. Mabuchi, E. S. Polzik, and H. J. Kimble,
	"Blue-light-induced infrared absorption in KNbO$_3$",
	J. Opt. Soc. Am. B {\bf 11}, 2023-2029 (1994).
\end{thebibliography}
\end{document}